\documentclass[aps,prd,preprint,groupedaddress,showpacs,nofootinbib]{revtex4}

\def\Vec#1{\mbox{\boldmath $#1$}}
\usepackage{graphicx}

\begin{document}


\title{Asymmetric neutrino emission due to neutrino-nucleon scatterings
in supernova magnetic fields}

\author{Shin'ichiro Ando}
\email[Email address: ]{ando@utap.phys.s.u-tokyo.ac.jp}
\affiliation{Department of Physics, School of Science, The University of
Tokyo, 7-3-1 Hongo, Bunkyo-ku, Tokyo 113-0033, Japan}

\date{\today}

\begin{abstract}
We derive the cross section of neutrino-nucleon scatterings in supernova
 magnetic fields, including weak-magnetism and recoil corrections.
Since the weak interaction violates the parity, the scattering cross
 section asymmetrically depends on the directions of the neutrino
 momenta to the magnetic field; the origin of pulsar kicks may be
 explained by the mechanism.
An asymmetric neutrino emission (a drift flux) due to neutrino-nucleon
 scatterings is absent at the leading level of $\mathcal O(\mu_BB/T)$,
 where $\mu_B$ is the nucleon magneton, $B$ is the magnetic field
 strength, and $T$ is the matter temperature at a neutrinosphere.
This is because at this level the drift flux of the neutrinos are
 exactly canceled by that of the antineutrinos.
Hence, the relevant asymmetry in the neutrino emission is suppressed by
 much smaller coefficient of $\mathcal O(\mu_BB/m)$, where $m$ is the
 nucleon mass; detailed form of the relevant drift flux is also derived
 from the scattering cross section, using a simple diffusion
 approximation.
It appears that the asymmetric neutrino emission is too small to induce
 the observed pulsar kicks.
However, we note the fact that the drift flux is proportional to the
 deviation of the neutrino distribution function from the value of
 thermal equilibrium at neutrinosphere.
Since the deviation can be large for non-electron neutrinos, it is
 expected that there occurs cancellation between the deviation and the
 small suppression factor of $\mathcal O(\mu_BB/m)$.
Using a simple parameterization, we show that the drift flux due to
 neutrino-nucleon scatterings may be comparable to the leading term due
 to beta processes with nucleons, which has been estimated to give a
 relevant kick velocity when the magnetic field is sufficiently strong
 as $10^{15}$--$10^{16}$ G.
\end{abstract}


\maketitle

\section{Introduction \label{sec:Introduction}}

A core-collapse supernova explosion is one of the most spectacular
events in astrophysics; 99\% of its gravitational binding energy is
released as neutrinos, while only 1\% as the kinetic energy of a shock
wave.
Therefore, neutrinos play an essential role in supernova explosions, and
their detection by ground-based large water \v Cherenkov detectors,
such as Super-Kamiokande and Sudbury Neutrino Observatory, would provide
valuable information on the nature of neutrinos as well as supernova
physics (see Ref. \citep{Ando03b}, and references therein).

Because of their dominance in supernova energetics, the neutrinos may
also give the solution to a long-standing astrophysical mystery, or
``pulsar kicks.''
Recent analyses of pulsar proper motion give a mean birth velocity
200--500 km s$^{-1}$ \citep{Lyne94,Lorimer97,Hansen97,Cordes98,
Arzoumanian02}, with possibly a significant population having velocity
$\agt 1000$ km s$^{-1}$.
These values are much greater than the velocities of their progenitors
($\sim 30$ km s$^{-1}$).
A natural explanation for such high velocities is that supernova
explosions are asymmetric, and provide kicks to nascent neutron stars.
In this paper, we are concerned with models in which the pulsar kicks
arise from {\it magnetic field induced asymmetry} in neutrino emissions
from proto-neutron stars.
We do not deal with another type of mechanism which relies on
hydrodynamical instabilities in the collapsed stellar core
\citep{Burrows92b,Burrows95a,Janka94,Janka96,Herant94}, leading to
asymmetric matter ejection and/or asymmetric neutrino emission.
Concerning the latter mechanism, several numerical results gave rather
negative results \citep{Janka94,Burrows96}.

There are many past studies which have noted that parity violation in
the weak interaction can lead to asymmetric neutrino emission in the
strongly magnetized neutron star matter (see,
e.g., Refs. \citep{Horowitz98,Arras99a,Arras99b}).
\citet{Arras99b} have shown that an asymmetric neutrino emission (a
drift flux) that may give the pulsar kicks is mainly induced by
charged-current interactions with nucleons, $\nu_en\to e^-p$ and
$\bar\nu_ep\to e^+n$, in magnetic fields.
On the other hand, the asymmetric emissions due to neutral-current
interactions, $\nu N\to\nu N$, are found to be irrelevant because their
contributions cancel between neutrinos and antineutrinos.
Further their finding is that the drift flux exists only when the
distribution deviates significantly from the value of thermal
equilibrium.
This result has invalidated the previous optimistic estimation that
sufficient asymmetry can be obtained by multiple scatterings of
neutrinos by nucleons, slightly polarized by the magnetic field of
moderate strength ($\sim 10^{12}$ G) \citep{Horowitz98}.
Their naive estimation has shown that the asymmetry parameter for the
$\nu_e,\bar\nu_e$ flux would be sufficient for the observed pulsar kick
velocity, if the magnetic field is very strong, $B\sim
10^{15}$--$10^{16}~{\rm G}$. 

In the present study, we focus on the effect of weak magnetism and
recoil corrections on neutrino-nucleon scatterings (particularly
$\nu_{\mu,\tau}$-$N$ and $\bar\nu_{\mu,\tau}$-$N$) in supernova magnetic
fields, which has not been considered in the previous publications such
as Ref. \citep{Arras99b}.
We expect that including these two corrections changes the previous
conclusions, which are listed above, as follows.
The weak magnetism correction reduces antineutrino-nucleon cross
sections compared to those of neutrino-nucleon scatterings
\citep{Horowitz02}.
For the reason, the cancellation of asymmetry between $\nu$ and
$\bar\nu$, which were shown in Ref. \citep{Arras99b}, does not occur at
$\mathcal O(k/m)$ level, where $k$ is the neutrino energy and $m$ is the
nucleon mass.
Although the drift flux due to $\nu$-$N$ scatterings is suppressed by a
small factor of $\mathcal O(k/m)$, this term may be as large as the term
due to charged-current interactions between electron (anti)neutrinos and
nucleons.
This is because the drift flux is proportional to the deviations of
distribution functions from the value of thermal equilibrium at a
neutrinosphere \citep{Arras99b}.\footnote{In this paper, we use a term
``neutrinosphere'' as that above which neutrinos freely stream out,
while by terms such as ``decoupling surface,'' we intend the surface
within which neutrinos are kept in thermal equilibrium.}
For $\mu$ and $\tau$ neutrinos, with which we are concerned in this
study, the transport opacity is primarily resulted from $\nu$-$N$
scattering, whereas energy exchange is due to elastic $\nu$-$e^-$
scattering whose cross section is substantially smaller.
As a consequence, the decoupling layer locates at much deeper region
than the neutrinosphere and there should be large optical depth for
the asymmetric flux to develop.
On the other hand, for $\nu_e$ and $\bar\nu_e$, the difference from
equilibrium at the neutrinosphere is considered to be very small owing
to dominant $\nu_en\to e^-p$ and $\bar\nu_ep\to e^+n$ reactions, which
act as very efficient energy-exchanging and thermalizing reactions.
Thus, although the drift flux due to $\nu_{\mu,\tau}$ and $\bar\nu_
{\mu,\tau}$ is suppressed by a small coefficient of $\mathcal O(k/m)$
compared with that due to $\nu_e$ and $\bar\nu_e$, we believe that the
large deviations from equilibrium for non-electron neutrinos can counter
the suppression factor.

Motivated by the above reasoning, we first derive the cross section for
$\nu$-$N$ scatterings in magnetic field by including the weak magnetism
and recoil corrections of scattered nucleons.
In general, it is quite difficult to evaluate.
However, in the limit that the nucleons are nondegenerate, we can
present a simple form of the differential cross section, and give a
diffusion equation which enables us to discuss qualitatively whether the
effect we are tackling is important or not.
We show, as a result of the diffusion equation and a simple
parameterization, that the drift flux on which we concentrate can
contribute at a considerable level for reasonable parameter choice.

Throughout this paper, we do not consider the effect of neutrino
oscillation, although recent atmospheric \citep{Fukuda99}, solar
\citep{Fukuda02a}, and reactor experiments \citep{Eguchi03} have shown
that the neutrinos have the nonzero masses and mixings.
This treatment is justified because the flavor mixing between electron
and the other flavor neutrinos is strongly suppressed by matter effect.
Roughly speaking, the efficient flavor conversion takes place at a
so-called resonance region, where a condition $n_e\simeq \Delta m^2/G_F
k$ is satisfied, however, at much deeper region $n_e\gg \Delta m^2/G_Fk$
matter induced neutrino oscillation is highly suppressed (see, e.g.,
Ref. \citep{Dighe00}).
With the recently inferred parameters $\Delta m^2$ by the oscillation
experiments, the resonance region locates at rather outer envelope such
as O+C or He layers; at a region around the neutrinosphere on which we
are focusing in this study, matter potential strongly prevents electron
(anti)neutrinos from mixing with the other flavor (anti)neutrinos.
In other words, at the sufficiently deep region in stars, electron
(anti)neutrinos propagate as the mass-eigenstates.
For $\mu$ and $\tau$ flavor (anti)neutrinos, they are found maximally
mixing with each other, however, it does not affect the supernova
dynamics such as pulsar kicks, because they can not be distinguished in
supernovae.
Thus, there is no need to worry about the effects of neutrino
oscillation on supernova dynamics.
Several years ago, \citet{Kusenko96} proposed a very interesting
mechanism that the pulsar kicks may be induced by neutrino oscillation.
Unfortunately as we have already noted above, this mechanism does not
work unless we introduce very heavy sterile neutrinos, because with the
experimentally inferred parameters neutrino oscillation is strongly
suppressed near the neutrinosphere.

The remainder of this paper is organized as follows.
In Sec. \ref{sec:Matrix element and differential cross section}, we
derive a general formula for the matrix element and the cross section of
the $\nu$-$N$ scatterings by considering the weak-magnetism and recoil
corrections.
A more concrete form of the cross section is derived in
Sec. \ref{sec:Nondegenerate nucleon limit}, based on the reasonable
assumption that the nucleons are highly nondegenerate.
In Sec. \ref{sec:Diffusion equation for a drift flux}, using a simple
diffusion approximation, the drift flux due to $\nu$-$N$ scatterings is
given in a very simple form, and we discuss the implications of the
diffusion equation in Sec. \ref{sec:Discussion and conclusion}.

\section{Matrix element and differential cross section \label{sec:Matrix
element and differential cross section}}

We consider neutrino-nucleon scatterings via the neutral-current
interaction in magnetic fields.
The usual $V-A$ current is modified when the weak magnetism correction
is included as
\begin{equation}
c_V\gamma^\mu + F_2\frac{i\sigma^{\mu\nu}q_\nu}{2m}
 - c_A\gamma^\mu\gamma^5,
\label{eq:corrected V-A}
\end{equation}
where $q^\mu=k^\mu -(k^\mu)^\prime$ is the momentum transfered to the
nucleon.
We label the neutrino momentum by $k$, and the nucleon momentum by $p$;
for the final state momenta, the symbol ${}^\prime$ is attached.
We summarize the values of couplings, $c_V,c_A$, and $F_2$ in Table
\ref{table:couplings}.

\begin{table}[htbp]
\caption{Coupling constants. Here $g_A\approx 1.26, \sin^2\theta_W
 \approx 0.23,\mu_p=1.793$, and $\mu_n=-1.913$. \label{table:couplings}}
\begin{tabular}{lccc} \hline \hline
 Reaction & $c_V$ & $c_A$ & $F_2$ \\ \hline
 $\nu p\to\nu p$ & $\frac{1}{2}-2\sin^2\theta_W$ & $\frac{g_A}{2}$
 & $\frac{1}{2}(\mu_p-\mu_n)-2\mu_p\sin^2\theta_W$ \\
 $\nu n\to\nu n$ & $-\frac{1}{2}$ & $-\frac{g_A}{2}$ 
 & $-\frac{1}{2}(\mu_p-\mu_n)-2\mu_n\sin^2\theta_W$ \\
 $\nu_en\to e^-p$ & $1$ & $g_A$ & $\mu_p-\mu_n$ \\
 $\bar\nu_ep\to e^+n$ & $1$ & $g_A$ & $\mu_p-\mu_n$
 \\ \hline \hline
\end{tabular}
\end{table}

Since the weak magnetism correction is at $k/m$ level, we must also
include the nucleon recoil correction of the same order.
The initial and final states of nucleon spinor can then be expressed by
\begin{eqnarray}
 u_N^s(p)\bar u_N^s(p) &=&
 \frac{1}{2}\left[(1-s\gamma^5\gamma^3)
 \left(1+\gamma^0+\frac{p_i}{m}\gamma^i\right)
 -s\frac{p_3}{m}\gamma^5(1+\gamma^0)
	 \right],\nonumber\\
 u_N^{s^\prime}(p^\prime)\bar u_N^{s^\prime}(p^\prime) &=&
 \frac{1}{2}\left[(1-s^\prime\gamma^5\gamma^3)
 \left(1+\gamma^0+\frac{p_i^\prime}{m}\gamma^i\right)
 -s^\prime\frac{p_3^\prime}{m}\gamma^5(1+\gamma^0)
	 \right],
\label{eq:recoiled spinor}
\end{eqnarray}
where $s,s^\prime=\pm 1$ specify the initial and final nucleon spins
relative to the magnetic field axis (3-axis).

With these corrections, the matrix element for $\nu$-$N$ scattering is
given, to the order of $k/m$, by
\begin{eqnarray}
|\mathcal M_{ss^\prime}(\Vec\Omega,\Vec\Omega^\prime,
 \Vec p,\Vec p^\prime)|^2&=&
 \left|\mathcal M_{ss^\prime}^{(0)}(\Vec\Omega,\Vec\Omega^\prime)
\right|^2
 +\delta|\mathcal M_{ss^\prime}(\Vec\Omega,\Vec\Omega^\prime,
 \Vec q)|^2
+\delta|\tilde\mathcal M_{ss^\prime}(\Vec\Omega,
  \Vec\Omega^\prime,\Vec P)|^2,
\label{eq:matrix element}
\end{eqnarray}
where $\Vec\Omega=\Vec k/k$ and $\Vec\Omega^\prime=\Vec
k^\prime/k^\prime$.
$\left|\mathcal M_{ss^\prime}^{(0)}(\Vec\Omega,\Vec\Omega^\prime)
\right|^2$ is the leading term already given in Ref. \citep{Arras99b}
by,
\begin{eqnarray}
\left|\mathcal M_{ss^\prime}^{(0)}(\Vec\Omega,\Vec\Omega^\prime)\right|
 ^2&=&\frac{G_F^2}{2}[c_V^2+3c_A^2+(c_V^2-c_A^2)\Vec\Omega\cdot\Vec
 \Omega^\prime+2c_A(c_V+c_A)(s\Vec\Omega+s^\prime\Vec\Omega^\prime)
 \cdot\Vec{\hat B}\nonumber\\
 &&{}+2c_A(c_V-c_A)(s\Vec\Omega^\prime+s^\prime\Vec\Omega)\cdot
  \Vec{\hat B}+ss^\prime\{(c_V^2-c_A^2)(1+\Vec\Omega\cdot\Vec\Omega
  ^\prime)\nonumber\\
 &&{}+4c_A^2\Vec\Omega\cdot\Vec{\hat B}\Vec\Omega^\prime
  \cdot\Vec{\hat B}\}],
\label{eq:matrix element at leading order}
\end{eqnarray}
$\delta |\mathcal M_{ss^\prime}(\Vec\Omega,\Vec\Omega^\prime,\Vec q)
|^2$ is the correction term at $\mathcal O(k/m)$ which depends on $\Vec
q(=\Vec p^\prime -\Vec p=\Vec k-\Vec k^\prime)$ as well as $\Vec\Omega$
and $\Vec\Omega^\prime$:
\begin{eqnarray}
\delta|\mathcal M_{ss^\prime}(\Vec\Omega,\Vec\Omega^\prime,\Vec q)|^2
 &=&\frac{G_F^2}{2m}[
 \pm 2c_A(c_V+F_2)\Vec q\cdot(\Vec\Omega-\Vec\Omega^\prime)+2c_A
 (c_V+F_2)(s-s^\prime)\Vec q\cdot\Vec{\hat B}\nonumber\\
 &&{}\pm2c_A(c_V+F_2)ss^\prime\Vec q\cdot\Vec{\hat B}
  (\Vec\Omega^\prime -\Vec\Omega)\cdot\Vec{\hat B}\nonumber\\
 &&{}+c_A
  (c_V+F_2)(s^\prime-s)(\Vec q\cdot\Vec\Omega\Vec
  \Omega^\prime\cdot\Vec{\hat B}+\Vec q\cdot\Vec\Omega^\prime
  \Vec\Omega\cdot\Vec{\hat B})\nonumber\\
 &&{}\mp c_V(c_V+F_2)(s+s^\prime)\{(\Vec\Omega^\prime\times
  \Vec\Omega)\times\Vec q\}\cdot\Vec{\hat B}],
\label{eq:matrix element depending on q}
\end{eqnarray}
and $\delta |\tilde\mathcal M_{ss^\prime}(\Vec\Omega,
\Vec\Omega^\prime,\Vec P)|^2$ is also the correction term at $\mathcal
O(k/m)$, which depends on $\Vec P(\equiv\Vec p+\Vec p^\prime)$:
\begin{eqnarray}
\delta |\tilde\mathcal M_{ss^\prime}(\Vec\Omega,\Vec\Omega^\prime,
 \Vec P)|^2&=&\frac{G_F^2}{2m}[
 -\{c_V^2+c_A^2+(c_V^2-c_A^2)ss^\prime\}\Vec P\cdot(\Vec\Omega
 +\Vec\Omega^\prime)-2c_Vc_A(s+s^\prime)\Vec P\cdot\Vec{\hat B}
 \nonumber\\&&{}
 -2c_A^2ss^\prime\Vec P\cdot\Vec{\hat B}(\Vec\Omega+\Vec\Omega
 ^\prime)\cdot\Vec{\hat B}-c_Vc_A(s+s^\prime)(\Vec P\cdot\Vec\Omega
 \Vec\Omega^\prime\cdot\Vec{\hat B}\nonumber\\
 &&{}+\Vec P\cdot\Vec\Omega^\prime
 \Vec\Omega\cdot\Vec{\hat B})\pm c_A (s^\prime -s)\{
 (\Vec\Omega^\prime\times\Vec\Omega)\times\Vec P\}\cdot\Vec{\hat B}].
\label{eq:matrix element depending on P}
\end{eqnarray}
Here, upper and lower signs represent the expressions for neutrinos and
antineutrinos, respectively; this notation is kept throughout this
paper.
Note that we can explicitly check the time-reversal symmetry for the
matrix element $|\mathcal M_{ss^\prime}(\Vec\Omega,\Vec\Omega^\prime,
\Vec p,\Vec p^\prime)|^2$, by exchanging all the initial and final state
quantities, i.e., $\Vec\Omega\leftrightarrow\Vec\Omega^\prime,\Vec
P\to\Vec P,\Vec q\to -\Vec q$, and $s\leftrightarrow s^\prime$.

The differential cross section (per unit volume) can be obtained from
the matrix element through the expression,
\begin{eqnarray}
\frac{d\Gamma}{dk^\prime d\Omega^\prime}&=&\frac{k^{\prime 2}}
 {(2\pi)^3}\sum_{ss^\prime}\int\frac{d^3p}{(2\pi)^3}
 \frac{d^3p^\prime}{(2\pi)^3}
 (2\pi)^4\delta^4(p+k-p^\prime -k^\prime)
  f_N(1-f_N^\prime)\nonumber\\
 &&{}\times|\mathcal M_{ss^\prime}(\Vec\Omega,\Vec\Omega^\prime,
  \Vec p,\Vec p^\prime)|^2,
\label{eq:differential cross section}
\end{eqnarray}
where $f_N\equiv f_N(E)$, $f_N^\prime\equiv f_N(E^\prime)$, $f_N(E)$ is
the nucleon distribution function with energy $E$ given by
\begin{equation}
f_N(E)=\frac{1}{\exp[(E-\mu_N)/T]+1},
\label{eq:nucleon distribution function}
\end{equation}
and $\mu_N$ is the nucleon chemical potential (excluding rest mass).
In evaluating the momentum integral, we neglect the Landau levels of
nucleons, and therefore the nucleon momentum is a well-defined quantity.
This is justified because many Landau levels are occupied for the
conditions in a protoneutron star, and the change in the available phase
space due to the Landau levels is negligible.
By substituting Eqs. (\ref{eq:matrix element at leading
order})--(\ref{eq:matrix element depending on P}) into
Eq. (\ref{eq:differential cross section}), the $\nu$-$N$ cross section
in the magnetic fields can be calculated to the order of $k/m$.
Equation (\ref{eq:differential cross section}) is written in more
convenient form as
\begin{eqnarray}
 \frac{d\Gamma}{dk^\prime d\Omega^\prime}&=&
  A_0(k,k^\prime,\mu^\prime)
  +\delta A_{\pm}(k,k^\prime,\mu^\prime)
  \Vec\Omega\cdot\Vec{\hat B}
  +\delta A_{\mp}(k,k^\prime,\mu^\prime)
  \Vec\Omega^\prime\cdot\Vec{\hat B}
  \nonumber\\
 &&{}+A_0^{WM}(k,k^\prime,\mu^\prime) 
  +\delta B^{WM}(k,k^\prime,\mu^\prime)
  \Vec\Omega\cdot\Vec{\hat B}
  +\delta C^{WM}(k,k^\prime,\mu^\prime)
  \Vec\Omega^\prime\cdot\Vec{\hat B},
\label{eq:reduced cross section for nondegenerate nucleons}
\end{eqnarray}
where $\mu^\prime =\Vec\Omega\cdot\Vec\Omega^\prime$.
$A_0$ and $\delta A_{\pm}$ are the terms which include the matrix
element $\left|\mathcal M_{ss^\prime}^{(0)}(\Vec\Omega,\Vec\Omega
^\prime)\right|^2$:
\begin{eqnarray}
A_0(k,k^\prime,\mu^\prime)&=&\frac{k^{\prime 2}}{(2\pi)^3}
 \sum_{s,s^\prime}\left|\mathcal M_{ss^\prime}^{(0)}
		   (\Vec\Omega,\Vec\Omega^\prime)\right|^2
 S_0(q_0,q),
\label{eq:A_0}\\
\delta A_{\pm}\Vec\Omega\cdot\Vec{\hat B}
 +\delta A_{\mp}\Vec\Omega^\prime\cdot\Vec{\hat B}
 &=&\frac{k^{\prime 2}}{(2\pi)^3}\sum_{s,s^\prime}
  \left|\mathcal M_{ss^\prime}^{(0)}
   (\Vec\Omega,\Vec\Omega^\prime)\right|^2\delta S_{ss^\prime}
  (q_0,q),
\label{eq:delta A}
\end{eqnarray}
where $S_0(q_0,q)$ and $\delta S_{ss^\prime}(q_0,q)$ can be derived
from the ``nuclear response function,'' $S_{ss^\prime}(q_0,q)=S_0(q_0,q)
+\delta S_{ss^\prime}(q_0,q)$, defined by
\begin{equation}
S_{ss^\prime}(q_0,q)=\int\frac{d^3p}{(2\pi)^3}
 \frac{d^3p^\prime}{(2\pi)^3}(2\pi)^4\delta^4
 (p+k-p^\prime-k^\prime)f_N(1-f_N^\prime).
\label{eq:nuclear response function}
\end{equation}
$S_0(q_0,q)$ is the leading term when $B=0$, and $\delta S_{ss^\prime}
(q_0,q)$ is the correction arising from nonzero $B$, which can be
simplified, for small $B$ compared to temperature ($\mu_BB/T\ll 1$), to
be 
\begin{eqnarray}
S_0(q_0,q)&=&\frac{m^2T}{2\pi q}\frac{1}{1-e^{-z}}
 \ln\left(\frac{1+e^{-x_0}}{1+e^{-x_0-z}}\right),
 \label{eq:S_0}\\
 \delta S_{ss^\prime}&=&-\frac{m^2T}{2\pi q}
  \frac{\delta x}{\left(e^{x_0}+1\right)
  \left(1+e^{-x_0-z}\right)},
  \label{eq:delta S}
\end{eqnarray}
with the parameters given by
\begin{eqnarray}
z&=&\frac{q_0}{T},~~x_0=\frac{(q_0-q^2/2m)^2}{4T(q^2/2m)}
 -\frac{\mu_N}{T},\nonumber\\
 \delta x&=&-\frac{\mu_BB}{2T}
  \left[\left(1+\frac{2mq_0}{q^2}\right)s
 +\left(1-\frac{2mq_0}{q^2}s^\prime\right)\right].
\label{eq:parameters}
\end{eqnarray}
The detailed derivations of these relations have been already discussed
in Ref. \citep{Arras99b}, and we refer the reader to the literature.
The terms including $A_0^{WM}$, $\delta B^{WM}$, and $\delta C^{WM}$,
which are smaller than $A_0$ and $\delta A_\pm$ by factor $k/m$, are
calculated by
\begin{eqnarray}
\lefteqn{A_0^{WM}+\delta B^{WM}\Vec\Omega\cdot\Vec{\hat B}
 +\delta C^{WM}\Vec\Omega^\prime\cdot\Vec{\hat B}}\nonumber\\
 &=&\frac{k^{\prime 2}}{(2\pi)^3}\sum_{s,s^\prime}
  \left[
  \delta |\mathcal M_{ss^\prime}(\Vec\Omega,\Vec\Omega^\prime,
  \Vec q)|^2\left\{S_0(q_0,q)+\delta S_{ss^\prime}
	   (q_0,q)\right\}\right.\nonumber\\
 &&{}\left.+\int\frac{d^3p}{(2\pi)^3}
\frac{d^3p^\prime}{(2\pi)^3}(2\pi)^4\delta^4
(p+k-p^\prime-k^\prime)f_N(1-f_N^\prime)\delta
|\tilde{\mathcal M}_{ss^\prime}(\Vec\Omega,\Vec\Omega^\prime,
\Vec P)|^2\right],
\label{eq:correction terms}
\end{eqnarray}
where in the second term, we cannot take the matrix element out of the
integral because it depends on $\Vec p$ and $\Vec p^\prime$, unlike that
in the first term which depends on $\Vec q=\Vec k-\Vec k^\prime$.

Equations (\ref{eq:A_0}), (\ref{eq:delta A}), and (\ref{eq:correction
terms}) are quite difficult to evaluate at this stage.
However, we can calculate it assuming that the nucleons are
nondegenerate, as we show in the next section.
In fact, this assumption appears to be appropriate, because the
asymmetric drift flux can develop only when the neutrino distribution
deviates from thermal equilibrium; this occurs in the regime where
nucleons are nondegenerate.

\section{Nondegenerate nucleon limit \label{sec:Nondegenerate nucleon
limit}}

For nondegenerate nucleons, the characteristic neutrino energy transfer
in each scattering is of order $q_0\sim k(T/m)^{1/2}\ll k$.
The cross section sharply peaks around $k^\prime =k$, and we can
evaluate $A_0,\delta A_{\pm},A_0^{WM},\delta B^{WM},$ and
$\delta C^{WM}$, in a series of the small parameter $(T/m)^{1/2}$.
We define the dimensionless quantities
\begin{equation}
\epsilon =\left[\frac{4(1-\mu^\prime)T}{m}\right]^{1/2},
 ~~u=\frac{k^\prime -k}{\epsilon k},
\label{eq:epsilon and u}
\end{equation}
where the range of $u$ is from $-1/\epsilon$ to $\infty$.
Using the expansions of the momentum integrals given in Appendix
\ref{sec:Derivation of differential cross section}, we
obtain the following expressions,
\begin{eqnarray}
 S_0(q_0,q)&=&\frac{m^2T}{2\pi q}n
  \left(\frac{2\pi^3}{m^3T^3}\right)^{\frac{1}{2}}
  \exp\left[-\frac{\left(q_0-q^2/2m\right)^2}
     {4Tq^2/(2m)}\right]
  \nonumber\\
 &\simeq&\frac{\pi^{1/2}n}{\epsilon k}
  e^{-u^2}\left[1-\frac{1}{2}
	   \left(1+\frac{k}{T}\right)\epsilon u
	   +\epsilon u^3
	   +\mathcal O(\epsilon^2)\right],
 \label{eq:S_0 for nondegenerate}\\
 A_0&=&\frac{k^{\prime 2}}{(2\pi)^3}2G_F^2
 \left[c_V^2+3c_A^2+(c_V^2-c_A^2)\mu^\prime\right]
 S_0(q_0,q),
 \label{eq:A_0 for nondegenerate}\\
 \delta A_{\pm}&=&\frac{k^{\prime 2}}{(2\pi)^3}4G_F^2c_A
  \frac{\mu_BB}{T}
  \left(c_V\pm 2c_A\frac{mq_0}{q^2}\right)S_0(q_0,q),
 \label{eq:delta A for nondegenerate}\\
 A_0^{WM}&=&-\frac{k^{\prime 2}}{(2\pi)^3}
  \frac{4G_F^2}{m}
  \left[(c_V^2+c_A^2)\frac{mq_0}{q^2}\Vec q\cdot 
   (\Vec\Omega +\Vec\Omega^\prime)
   \pm c_A(c_V+F_2)\Vec q\cdot(\Vec\Omega^\prime
   -\Vec\Omega)\right]\nonumber\\
 &&{}\times S_0(q_0,q),
 \label{eq:A_WM for nondegenerate}\\
 \delta B^{WM}&=&\frac{k^{\prime 2}}{(2\pi)^3}
  \frac{2G_F^2}{m}\frac{\mu_BB}{T}
  \left[4c_AF_2\frac{mq_0}{q^2}k-2c_A(2c_V+F_2)
   \frac{mq_0}{q^2}\Vec q\cdot\Vec\Omega^\prime\right.
   \nonumber\\
 &&{}\left.\pm\left\{4c_A^2\frac{mT}{q^2}
 \left(1-\frac{mq_0^2}{Tq^2}\right)
 -c_V(c_V+F_2)\right\}\Vec q
 \cdot\Vec\Omega^\prime\right]S_0(q_0,q),
 \label{eq:delta B_WM for nondegenerate}\\
 \delta C^{WM}&=&\frac{k^{\prime 2}}{(2\pi)^3}
  \frac{2G_F^2}{m}\frac{\mu_BB}{T}
  \left[-4c_AF_2\frac{mq_0}{q^2}k^\prime-2c_A(2c_V+F_2)
   \frac{mq_0}{q^2}\Vec q\cdot\Vec\Omega\right.
   \nonumber\\
 &&{}\left.\mp\left\{4c_A^2\frac{mT}{q^2}
 \left(1-\frac{mq_0^2}{Tq^2}\right)
 -c_V(c_V+F_2)\right\}\Vec q
 \cdot\Vec\Omega\right]S_0(q_0,q),
 \label{eq:delta C_WM for nondegenerate}
\end{eqnarray}
where all of them are expressed by $u$ and $\epsilon$ instead of $q_0$
and $q$, through the relations such as $q_0=k-k^\prime=-\epsilon uk$ and
$\Vec q=\Vec k-\Vec k^\prime=k\Vec\Omega -k(1+\epsilon u)\Vec\Omega
^\prime$.
In deriving the expressions above, we have used the equation $e^{\mu_
N/T}=n(2\pi^3/m^3T^3)^{1/2}$ that is valid when $B=0$, to relate the
nucleon chemical potential $\mu_N$ to its number density $n$ (the
corrections due to finite $B$ are of order $B^2$).
These expressions are valid under the conditions $T\ll m$, $k\ll
(mT)^{1/2}$, $\mu_BB\ll T$, and $k\agt k_{\rm min}=\mu_BB(m/T)
^{1/2}\simeq 10^{-2}|g|B_{14}T^{-1/2}$ MeV.
All these conditions are satisfied for the case of interest in our
study.
Unlike the previous publication \citep{Arras99b}, we must keep in the
bracket of $S_0$ to $\mathcal O(\epsilon^4)$ terms although we did not
explicitly show them in Eq. (\ref{eq:S_0 for nondegenerate}).
This is because they are necessary for the evaluation of the drift flux
which is suppressed at $\mathcal O(k/m)$ level compared with the leading
term.

Thus, we have derived the useful form of the differential cross section
of neutrino-nucleon scatterings, which enables us to discuss the parity
violating effect of the neutral-current weak interaction related to
pulsar kicks.
At first sight, however, these equations are still too complicated to
obtain an intuitive implication for the asymmetric neutrino emission in
the supernova magnetic field.
Therefore, in the remainder of this paper, we derive a simple diffusion
equation with which we can see how large drift flux is obtained.
More delicate treatment using numerical simulations is beyond the scope
of this study; it is slated for the future work.

\section{Diffusion equation for a drift flux \label{sec:Diffusion
equation for a drift flux}}

The Boltzmann equation for neutrino transport is written in the form
\begin{equation}
 \frac{\partial f_\nu(\Vec k)}{\partial t}
  +\Vec\Omega\cdot\nabla f_\nu(\Vec k)
  =\left[\frac{\partial f_\nu(\Vec k)}{\partial t}
  \right]_{\rm sc},
\label{eq:Boltzmann equation}
\end{equation}
where we include the collision term due to scattering alone; effects of 
absorption are not considered in this section.
The collision term in Eq. (\ref{eq:Boltzmann equation}) can be written
by the following integral \citep{Arras99b}:
\begin{equation}
\left[\frac{\partial f_\nu(\Vec k)}{\partial t}\right]_{\rm sc}
 =\int_0^\infty dk^\prime\int d\Omega^\prime
 \frac{d\Gamma}{dk^\prime d\Omega^\prime}
 \left[C(k,k^\prime)\delta f_\nu^\prime
  +D(k,k^\prime)\delta f_\nu\right],
\label{eq:collision term due to scattering}
\end{equation}
where $\delta f_\nu$ and $\delta f_\nu^\prime$ denote the deviation of
neutrino distribution function from the value of equilibrium, i.e.,
$\delta f_\nu =f_\nu -f_\nu^{(0)}$.
The coefficients of $\delta f_\nu$ and $\delta f_\nu^\prime$ are
\begin{eqnarray}
 C(k,k^\prime)&=&e^{-q_0/T}(1-f_\nu^{(0)})+f_\nu^{(0)},
 \label{eq:C}\\
 D(k,k^\prime)&=&-\left[e^{-q_0/T}f_\nu^{(0)\prime}
		 +1-f_\nu^{(0)\prime}\right].
 \label{eq:D}
\end{eqnarray}
The nonlinear terms which are proportional to $\delta f_\nu\delta f_\nu^
\prime$ were dropped since we consider the regime where the deviation
from thermal equilibrium $\delta f_\nu$ is relatively small (the regime
where the diffusion approximation is valid).

For the purpose of examining the macroscopic consequence of the
asymmetric cross section, we expand the neutrino distribution function
in spherical harmonics up to dipole order as
\begin{equation}
\delta f_\nu(\Vec k)=g(k)+3\Vec\Omega\cdot\Vec h(k),
\label{eq:harmonic expansion of f_nu}
\end{equation}
where $g(k)$ is the spherically symmetric deviation from the
equilibrium, and $\Vec h(k)$ represents the dipole deviation which leads
to the flux.
These quantities, $g(k)$ and $\Vec h(k)$ are related to the energy
density and the flux through the relations
\begin{eqnarray}
 U_\nu(k)&=&\int\frac{k^2d\Omega}{(2\pi)^3}kf_\nu
  =\frac{k^3}{2\pi^2}\left[f_\nu^{(0)}(k)+g(k)\right]
  =U_\nu^{\rm FD}(k)+\frac{k^3}{2\pi^2}g(k),
  \label{eq:g to energy density}\\
 \Vec F_\nu(k)&=&\int\frac{k^2d\Omega}
  {(2\pi)^3}k\Vec\Omega f_\nu
  =\frac{k^3}{2\pi^2}\Vec h(k),
  \label{eq:h to flux}
\end{eqnarray}
where the quantity with superscript FD represents the value in the case
of thermal equilibrium.

The first moment of Eq. (\ref{eq:collision term due to scattering})
yields
\begin{equation}
 \int\frac{d\Omega}{4\pi}\Vec\Omega
 \left[\frac{\partial f_\nu(\Vec k)}{\partial t}\right]_{\rm sc}
 =\Vec V_0\pm\delta\Vec V_\pm^B
 +\delta\Vec V_0^{WM}\pm\delta\Vec V_\pm^{WM}
 +\delta\Vec V_0^{B,WM}\pm\delta\Vec V_\pm^{B,WM},
\end{equation}
where the subscript ``$0$'' of $\Vec V$ and $\delta\Vec V$ represents
that the terms are the same for neutrinos and antineutrinos, while
``$\pm$'' that the sign of each term is different for neutrinos and
antineutrinos.
The term $\Vec V_0$ contributes at the leading level, whereas the
quantities with superscript ``$B$,'' ``$WM$,'' and ``$B,WM$'' are
suppressed by factors of $\mathcal O(\mu_BB/T)$, $\mathcal O(T/m)$, and
$\mathcal O(\mu_BB/m)$, respectively.
As a result of lengthy calculations (details are given in
Ref. \citep{Arras99b}), each term is written as
\begin{eqnarray}
 \Vec V_0&=&-\frac{2G_F^2k^2}{3\pi}n(c_V^2+5c_A^2)\Vec h,
  \label{eq:V_0}\\
 \delta\Vec V_\pm^B&=&-\frac{4G_F^2k^2}{3\pi}n\frac{\mu_BB}{T}
  c_A^2\left[\left(1-2f_\nu^{(0)}\right)g
      +T\frac{\partial g}{\partial k}\right]\Vec{\hat B},
  \label{eq:delta V_B}\\
 \delta \Vec V_0^{WM}&=&-\frac{2G_F^2k^2}{3\pi}n\frac{k}{m}
  \left[\left\{-3(c_V^2+3c_A^2)\left(1-2f_\nu^{(0)}\right)
	 +5(c_V^2+9c_A^2)\frac{T}{k}
	 \right.\right.\nonumber\\
 &&{}\left.\left.+2(c_V^2+5c_A^2)k\frac{\partial f_\nu^{(0)}}
	 {\partial k}\right\}\Vec h
	 +2c_A^2\left(1-2f_\nu^{(0)}
		 +\frac{6T}{k}\right)
	 k\frac{\partial\Vec h}{\partial k}
	 +2c_A^2kT\frac{\partial^2\Vec h}
	 {\partial k^2}\right],
 \label{eq:delta V_WM_0}\\
 \delta \Vec V_\pm^{WM}&=&-\frac{16G_F^2k^2}{3\pi}n\frac{k}{m}
  c_A(c_V+F_2)\Vec h,
  \label{eq:delta V_WM_pm}\\
 \delta \Vec V_0^{B,WM}&=&-\frac{4G_F^2k^2}{9\pi}n
  \frac{\mu_BB}{T}\frac{k}{m}c_A
  \left[\left\{(-2c_V+5F_2)\left(1-2f_\nu^{(0)}\right)
	 +2c_Vk\frac{\partial f_\nu^{(0)}}
	 {\partial k}\right\}g
	 \right.\nonumber\\
 &&{}\left.-\left\{c_V\left(1-2f_\nu^{(0)}\right)
	 +(2c_V-5F_2)\frac{T}{k}\right\}
	 k\frac{\partial g}{\partial k}
	 -c_VkT\frac{\partial^2 g}
	 {\partial k^2}\right]\Vec{\hat B},
 \label{eq:delta V_BWM_0}\\
 \delta \Vec V_\pm^{B,WM}&=&
  -\frac{4G_F^2k^2}{9\pi}n\frac{\mu_BB}{m}c_A^2
  \left[\left\{69\left(1-2f_\nu^{(0)}\right)
	 -68k\frac{\partial f_\nu^{(0)}}{\partial k}
 \right.\right.\nonumber\\	 
 &&{}\left.-12k^2\frac{\partial^2 f_\nu^{(0)}}
	    {\partial k^2}\right\}g
	    +\left\{34\left(1-2f_\nu^{(0)}\right)+\frac{2k}{T}
	    +\frac{69T}{k}\right\}
	    k\frac{\partial g}
	    {\partial k}\nonumber\\
 &&{}\left.+2\left\{3\left(1-2f_\nu^{(0)}\right)
	       +\frac{17T}{k}\right\}
	       k^2\frac{\partial^2 g}{\partial k^2}
	       +4k^2T\frac{\partial^3 g}
	       {\partial k^3}\right]\Vec{\hat B}.
 \label{eq:delta V_BWM_pm}
\end{eqnarray}
Since the sign of the $\mathcal O(\mu_BB/T)$ term is opposite for
neutrinos and antineutrinos, there occurs cancellation of the drift flux
at this level.
Hence, the leading drift flux due to $\nu$-$N$ scatterings is suppressed
by a very small coefficient of $\mathcal O(\mu_BB/m)$.

When we discuss the terms which depend on quantities such as $\partial
\Vec h/\partial k,\partial^2 \Vec h/\partial k^2$, we use the lowest
order expression for $\Vec h$, i.e.,
\begin{eqnarray}
 \Vec h(k)&\simeq&-\frac{1}{3\kappa_0}\nabla
 \left(f_\nu^{(0)}(k)+g(k)\right),
 \label{eq:lowest}\\
 \kappa_0&=&\frac{2G_F^2k^2}{3\pi}n(c_V^2+5c_A^2).
  \label{eq:kappa}
\end{eqnarray}
In this and the following expressions for the diffusion equations, we
drop the time derivative term; it corresponds to a rapid redistribution
of matter temperature, whose time scale is of order the mean free path
divided by the speed of light $c$, much smaller than neutrino diffusion
time of the star.
In addition, we assume that all the quantities except for $\Vec h$, such
as $g,f_\nu^{(0)}$, are the same for neutrinos and antineutrinos.
All these assumptions are considered to be quite applicable, because the
discrepancy caused by them is further suppressed to the extent in which
we are not interested.
With the above assumptions, we take the first moment of the Boltzmann
equation (\ref{eq:Boltzmann equation}), and obtain the explicit
expression for the neutrino plus antineutrino flux $\Vec F_\nu+\Vec
F_{\bar\nu}$:
\begin{eqnarray}
 \Vec F_\nu (k)+\Vec F_{\bar\nu}(k)&=&
  -\frac{2}{3\kappa_0}\left(1-\delta^{(1)}\right)
  \nabla U_\nu (k)
  \nonumber\\
 &&{}+\delta^{(2)}
  \left[\left(1-2f_\nu^{(0)}+\frac{6T}{k}\right)
   k\frac{\partial}{\partial k}
   \left(\frac{\nabla U_\nu (k)}{\kappa_0}\right)
   +kT\frac{\partial^2}{\partial k^2}
   \left(\frac{\nabla U_\nu (k)}{\kappa_0}\right)\right]
   \nonumber\\
 &&{}+\left[\epsilon^{(1)}\delta U_\nu (k)
       +\epsilon^{(2)}k\frac{\partial}{\partial k}
       \delta U_\nu (k)+\epsilon^{(3)}k^2
       \frac{\partial^2}{\partial k^2}
       \delta U_\nu (k)\right]\Vec{\hat B},
 \label{eq:flux}
\end{eqnarray}
where $\delta^{(1)}$ and $\delta^{(2)}$ are small numbers of $\mathcal
O(k/m)$, and $\epsilon^{(1)}$, $\epsilon^{(2)}$, and $\epsilon^{(3)}$
are those of $\mathcal O(\mu_BB/m)$; they are given in concrete forms by
\begin{eqnarray}
 \delta^{(1)}&=&\frac{k}{m}\frac{1}{c_V^2+5c_A^2}
  \left[-3\left(c_V^2+3c_A^2\right)\left(1-2f_\nu^{(0)}\right)
   -5\left(c_V^2+9c_A^2\right)\frac{T}{k}
 \right.\nonumber\\
 &&{}\left.
      +2\left(c_V^2+5c_A^2\right)k\frac{\partial f_\nu^{(0)}}
      {\partial k}\right],
 \label{eq:delta 1}\\
 \delta^{(2)}&=&\frac{4k}{3m}\frac{c_A^2}{c_V^2+5c_A^2},
  \label{eq:delta 2}\\
 \epsilon^{(1)}&=&\frac{4k}{3m}\frac{\mu_BB}{T}
  \frac{c_A}{\left(c_V^2+5c_A^2\right)^2}
  \left[\left\{2c_V\left(c_V^2+17c_A^2\right)
	 -F_2\left(5c_V^2+c_A^2\right)\right\}
 \left(1-2f_\nu^{(0)}\right)
 \right.\nonumber\\
 &&{}\left.-2c_V\left(c_V^2+5c_A^2\right)
 k\frac{\partial f_\nu^{(0)}}{\partial k}\right],
  \label{eq:epsilon 1}\\
 \epsilon^{(2)}&=&\frac{4k}{3m}\frac{\mu_BB}{T}
  \frac{c_A}{\left(c_V^2+5c_A^2\right)^2}
  \left[c_V^3\left(1+\frac{2T}{k}\right)
   +c_Vc_A^2\left(5+\frac{34T}{k}\right)
   \right.\nonumber\\
 &&{}\left.-2c_V\left(c_V^2+5c_A^2\right)f_\nu^{(0)}
      -F_2\left(5c_V^2+c_A^2\right)\frac{T}{k}\right],
      \label{eq:epsilon 2}\\
 \epsilon^{(3)}&=&\frac{4\mu_BB}{3m}
  \frac{c_Vc_A}{c_V^2+5c_A^2}.
  \label{eq:epsilon 3}
\end{eqnarray}
As expected and already pointed out by \citet{Arras99b}, the drift
flux, which is along the direction of the magnetic field $\Vec B$, is
absent at order $\mathcal O(\mu_BB/T)$; the leading term of the relevant
drift flux is at $\mathcal O(\mu_BB/m)$ as shown in
Eq. (\ref{eq:flux}).

\section{Discussion and conclusion \label{sec:Discussion and
conclusion}}

Neutrinos carry away almost all the energy released by gravitational
collapse of a supernova core, $\approx 3\times10^{53}$ erg, which is
$\sim 100$ times the momentum associated with the spatial motion of
pulsars.
Therefore, 1\% anisotropy in the momentum distribution of the outgoing
neutrinos would suffice to account for the observed kick velocities.
From Eq. (\ref{eq:flux}), the kick due to $\nu$-$N$ scatterings is
characterized by the fractional momentum asymmetry factor
\begin{eqnarray}
 \alpha_{\rm sc}&=&
  \frac{\int_0^\infty dk\int R_\nu^2(k)d\Omega
  [\Vec F_\nu(k)+\Vec F_{\bar\nu}(k)]\cdot\Vec{\hat B}}
  {\int_0^\infty dk\int R_\nu^2(k)d\Omega
  [\Vec F_\nu(k)+\Vec F_{\bar\nu}(k)]\cdot\Vec{\hat r}}
  \nonumber\\
 &\simeq&\frac{\int_0^\infty dkR^2_\nu(k)
  \left[\epsilon^{(1)}\delta U_\nu(k)
   +\epsilon^{(2)}k\frac{\partial}{\partial k}\delta U_\nu(k)
   +\epsilon^{(3)}k^2\frac{\partial^2}
   {\partial k^2}\delta U_\nu(k)\right]}
  {\int_0^\infty dkR^2_\nu(k)\frac{2}{3\kappa_0}
  \left|\frac{dU_\nu(k)}{dr}\right|_{\nu~{\rm sph.}}},
  \label{eq:asymmetry factor}
\end{eqnarray}
where $R_\nu(k)$ represents a radius of the neutrinosphere of energy
$k$.
To evaluate the value of $\alpha_{\rm sc}$ we must be careful, because
the numerator of Eq. (\ref{eq:asymmetry factor}) contains $\delta U_\nu$
which is difficult to estimate without numerical simulations.
However, we believe, for non-electron neutrinos and antineutrinos this
deviation $\delta U_\nu$ can be large, which may cancel the small
factors $\epsilon^{(i)}$.
This is because of the fact as follows.
The electron neutrinos and antineutrinos are kept in thermal equilibrium
slightly within the neutrino sphere by beta processes, $\nu_en
\leftrightarrow e^-p$ and $\bar\nu_ep\leftrightarrow e^+n$; leading very
little difference from thermal equilibrium $\delta U_{\nu_e,\bar\nu_e}$
at the neutrinosphere.
In contrast, for $\nu_{\mu,\tau}$ and $\bar\nu_{\mu,\tau}$ the main
source for the opacity is $\nu N\leftrightarrow\nu N$ reactions.
However, since the nucleon mass is much larger than the relevant
temperatures, energy exchange between neutrinos and nucleons is
inefficient.
Thus, the decoupling surface, within which neutrinos are kept in thermal
equilibrium by the subdominant energy exchange processes such as $NN
\leftrightarrow NN\nu\bar\nu,e^+e^-\leftrightarrow\nu\bar\nu,$ and $\nu
e\leftrightarrow \nu e$, locates at much deeper region than the
neutrinosphere; it should lead rather large deviation $\delta
U_\nu(k)$.

While a detailed study using numerical calculations is not given in the
present paper, we simply compare our new drift term in
Eq. (\ref{eq:flux}) with the previously derived drift term due to
absorption \citep{Arras99b}.
According to \citet{Arras99b}, the drift flux due to the absorption
$\nu_en\to e^-p$ is described by
\begin{equation}
 \Vec F_{\nu_e}^{\rm drift}
 =-\frac{\kappa_0^{*({\rm abs})}}{3\kappa_0^{({\rm tot})}}
 \epsilon_{\rm abs}\delta U_{\nu_e}(k)\Vec{\hat B},
 \label{eq:drift flux due to absorption}
\end{equation}
where $\kappa_0^{({\rm tot})}=\kappa_0^{*({\rm abs})}+\kappa_0$,
$\kappa_0^{*({\rm abs})}=\kappa_0^{({\rm abs})}[1+e^{(\mu_\nu-k)/T}]$,
with
\begin{eqnarray}
 \kappa_0^{({\rm abs})}&=&\frac{G_F^2}{\pi}(k+m_n-m_p)^2
  n_n(c_V^2+3c_A^2)[1-f_e(k+m_n-m_p)],
  \label{eq:kappa_0 for absorption}\\
 \epsilon_{\rm abs}&=&\frac{1}{2}\frac{eB}{(k+m_n-m_p)^2}
  \frac{c_V^2-c_A^2}{c_V^2+3c_A^2}
  +2\frac{c_A(c_A+c_V)}{c_V^2+3c_A^2}
  \frac{\mu_{Bn}B}{T}
  \nonumber\\
 &&{}-\frac{T}{k+m_n-m_p}
  \left[1+\frac{k+m_n-m_p}{T}f_e(k+m_n-m_p)\right]
  \nonumber\\
 &&{}\times\left[2\frac{c_A(c_A+c_V)}{c_V^2+3c_A^2}
	    \frac{\mu_{Bn}B}{T}+2\frac{c_A(c_A-c_V)}{c_V^2+3c_A^2}
	    \frac{\mu_{Bp}B}{T}\right].
\end{eqnarray}
In order to compare the difference between the drift flux due to
$\nu$-$N$ scatterings and that due to absorption, we define the quantity
$R_{\rm sc/abs}$ by
\begin{equation}
 R_{\rm sc/abs}=\left|\frac{2\int_0^\infty dk
  \left[\epsilon^{(1)}\delta U_{\nu_\mu}(k)
   +\epsilon^{(2)}k\frac{\partial}{\partial k}\delta U_{\nu_\mu}(k)
   +\epsilon^{(3)}k^2\frac{\partial^2}
   {\partial k^2}\delta U_{\nu_\mu}(k)\right]}
  {\int_0^\infty dk
  \frac{\kappa_0^{*({\rm abs})}}
  {3\kappa_0^{({\rm tot})}}\epsilon_{\rm abs}
  \delta U_{\nu_e}(k)}\right|,
\end{equation}
which is the ratio of the asymmetric fluxes due to the scattering and
the absorption.
The factor 2 in the numerator represents that the drift flux due to
scatterings mainly comes from two-flavor ($\mu$ and $\tau$) neutrinos,
whereas that due to absorptions is only related to electron neutrinos.
More precisely, $R^2_\nu(k)$ must be included in the integral, but for
simplicity we neglect it.
In order to evaluate $R_{\rm sc/abs}$, we need a further assumption
concerning the energy densities $U_{\nu_e}$ and $U_{\nu_\mu}$.
Here we parameterize them using the ``effective temperature'' $T_e^{\rm
eff}$ and $T_\mu^{\rm eff}$ as follows:
\begin{eqnarray}
 U_{\nu_e}(k)&=&\frac{k^3}{2\pi^2}
  \frac{1}{e^{k/T_e^{\rm eff}}+1},
  \label{eq:parametrize for U_nue}\\
 U_{\nu_\mu}(k)&=&\frac{k^3}{2\pi^2}
  \frac{1}{e^{k/T_\mu^{\rm eff}}+1},
  \label{eq:parametrize for U_numu}
\end{eqnarray}
where for further simplicity, $T^{\rm eff}_e$ and $T^{\rm eff}_\mu$ are
assumed to be independent of $k$.
As the other parameters at the neutrinosphere, we take $T=3$ MeV,
$\rho=10^{12}~\mathrm{g~cm^{-3}}$, and $Y_e=0.3$, and the reactions
(scatterings/absorptions) with neutrons alone are considered.

\begin{figure}[htbp]
\begin{center}
\includegraphics[width=12cm]{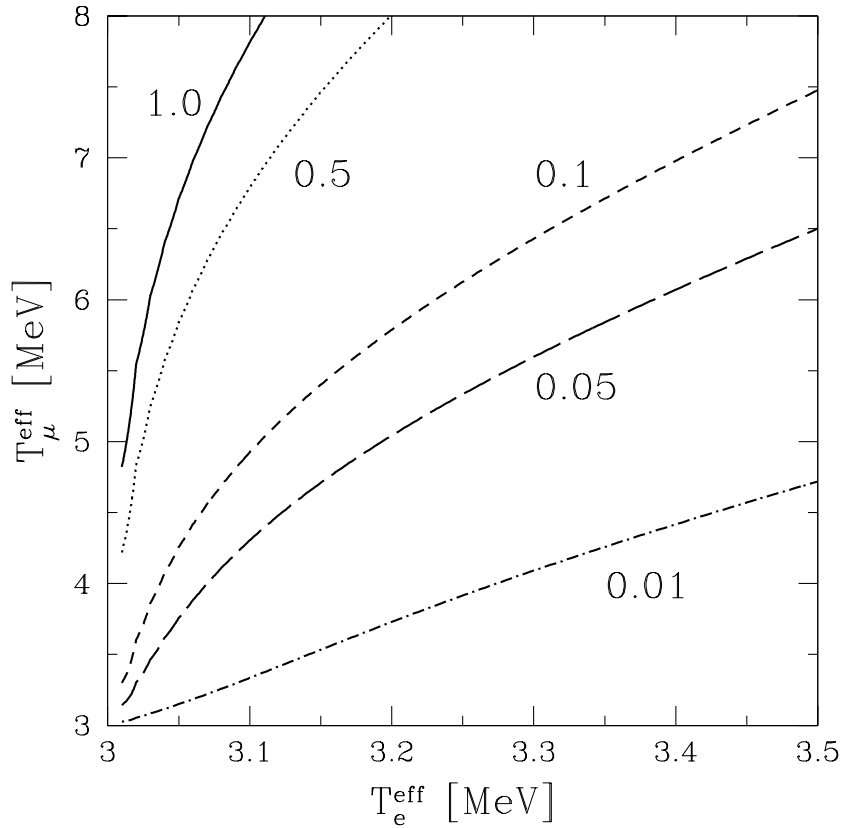}
\caption{Contour map for the value of $R_{\rm sc/abs}$ projected against
 $T_e^{\rm eff}$ and $T_\mu^{\rm eff}$. \label{fig:ratio}}
\end{center}
\end{figure}

Figure \ref{fig:ratio} shows a contour map for the value of $R_{\rm
sc/abs}$, which is projected against $T_e^{\rm eff}$ and $T_\mu^{\rm
eff}$.
As discussed above, we believe that the value of $T_e^{\rm eff}$ is very
close to the temperature at the neutrinosphere (3 MeV), on the other
hand, we take $T_\mu^{\rm eff}$ to the extremely high value, 8 MeV,
though it may be unlikely to be realized.\footnote{Of course, taking too
high value for the effective temperature violates the diffusion
approximation, but we believe that it still gives some useful
implications.}
This figure clearly shows that a significant part of the drift flux can
be attributed to $\nu_{\mu,\tau}$-$N$ scatterings, for a considerable
area in parameter space of $(T_e^{\rm eff},T_\mu^{\rm eff})$.
In particular, they are comparable to the effect by the charged-current
interactions, when $T_e^{\rm eff}$ is extremely close to the temperature
at the neutrinosphere, and this situation may be realized actually.
Thus, we cannot neglect the effect of $\nu$-$N$ scatterings even though
at the leading order it is canceled between neutrinos and antineutrinos;
it may give considerable contribution to the asymmetric neutrino
emission, in addition to the charged-current interactions.
Whether we can actually obtain the observed pulsar kicks from this
mechanism is beyond the scope of this study, and it is a subject for the
future numerical study.

In this paper, we have derived the cross sections of $\nu$-$N$
scatterings in strong magnetic fields, including weak-magnetism and
recoil corrections.
Since neutrino interactions are described by theory of the weak
interaction which violate the parity, the asymmetric neutrino emission
(drift flux) is induced, and it may give the origin of pulsar kicks.
The drift flux due to $\nu$-$N$ scatterings is suppressed at $\mathcal
O(\mu_BB/m)$ level, because the leading contribution is canceled between
neutrinos and antineutrinos.
However, since the drift flux is proportional to the deviation of
neutrino energy density from the value of thermal equilibrium $\delta
U_\nu$ and it can be large for non-electron neutrinos, it can cancel the
small suppression factor of $\mathcal O(\mu_BB/m)$.
We have shown that the drift flux due to $\nu$-$N$ scatterings may be
comparable to the leading term due to beta processes with nucleons.
It is expected to give the non-negligible effect for inducing relevant
pulsar kicks.

\appendix

\section{Derivation of differential cross section \label{sec:Derivation
of differential cross section}}

In this section, we derive differential cross section, Eqs. (\ref{eq:S_0
for nondegenerate})--(\ref{eq:delta C_WM for nondegenerate}), from more
general form, Eqs. (\ref{eq:A_0}), (\ref{eq:delta A}), and
(\ref{eq:correction terms}), in the case of nondegenerate nucleons.

We do not give the discussion for deriving the detailed form for nucleon
response function $S_{ss^\prime}$, because it has been already discussed
in Ref. \citep{Arras99b}, and we basically follow their method in
deriving the remaining part.
Our interest here is concerned with the momentum integral which includes
$\Vec P$, i.e.,
\begin{equation}
 \int\frac{d^3p}{(2\pi)^3}\frac{d^3p^\prime}{(2\pi)^3}
  (2\pi)^4\delta^4(p+k-p^\prime-k^\prime)
  f_N(1-f_N^\prime)\Vec P,
  \label{eq:P integral}
\end{equation}
which appears in the second term in the right hand side of
Eq. (\ref{eq:correction terms}).
For the term including $\Vec p$, we first use $d^3p^\prime$ to integrate
over $\delta^3(\Vec p+\Vec q-\Vec p^\prime)$ and then integrate over the
azimuthal angle for $\Vec p$, resulting in
\begin{equation}
\frac{\Vec q}{2\pi q}\int_0^\infty dpp^3\int_{-1}^1d\mu
 ~\mu\delta(q_0+E-E^\prime)f_N(E)\left[1-f_N(E^\prime)\right],
 \label{eq:after azimuthal integration}
\end{equation}
with $\mu=\Vec p\cdot\Vec q/pq$ and $E^\prime=-\mu_BBs^\prime+(\Vec p+
\Vec q)^2/2m=-\mu_BBs^\prime+(p^2+q^2+2pq\mu)/2m$.
Care must be taken to correctly integrate over the energy-conservation
delta function with coordinate $\mu$; the integral is only nonzero if
the argument of the delta function is zero for $\mu\in(-1,1)$.
This condition, $\mu^2\le 1$ is rewritten in the following expression
for momentum $p$ 
\begin{equation}
 p^2\ge p^2_{\rm min}=
  \left[\frac{q_0-q^2/2m-\mu_BB(s-s^\prime)}
   {q/m}\right]^2.
  \label{eq:integral condition}
\end{equation}
Here we change variables from $p$ to $E=-s\mu_BB+p^2/2m$ in the
remaining integral; it gives
\begin{eqnarray}
 \lefteqn{\frac{m^3\Vec q}{2\pi q^3}
  \int_{E_{\rm min}}^\infty dE
  \left[q_0-\frac{q^2}{2m}-\mu_BB(s-s^\prime)\right]
  f_N(E)[1-f_N(E+q_0)]}\nonumber\\
  &=&\frac{m\Vec q}{q^2}
  \left[q_0-\frac{q^2}{2m}-\mu_BB(s-s^\prime)\right]
  S_{ss^\prime}(q_0,q),
  \label{eq:final p integral}\\
 E_{\rm min}&=&-\mu_BBs+\frac{[q_0+q^2/2m-\mu_BB(s-s^\prime)]^2}
  {4(q^2/2m)}.
  \label{eq:E_min}
\end{eqnarray}
Note that the last form of Eq. (\ref{eq:final p integral}) contains the
integral which is also obtained from the integration of nucleon response
function $S_{ss^\prime}(q_0,q)$.
With the similar manner, the term of Eq. (\ref{eq:P integral}) which
includes $\Vec p^\prime$ is derived in the following form:
\begin{eqnarray}
 \lefteqn{\frac{m^3\Vec q}{2\pi q^3}
  \int_{E_{\rm min}^\prime}^\infty dE^\prime
  \left[q_0+\frac{q^2}{2m}-\mu_BB(s-s^\prime)\right]
  f_N(E^\prime-q_0)\left[1-f_N(E^\prime)\right]}\nonumber\\
  &=&\frac{m\Vec q}{q^2}\left[q_0+\frac{q^2}{2m}
		 -\mu_BB(s-s^\prime)\right]
  \tilde S_{ss^\prime}(q_0,q),
  \label{eq:final p' integral}
\end{eqnarray}
where
\begin{eqnarray}
 \tilde S_{ss^\prime}(q_0,q)&=&\frac{m^2}{2\pi q}
  \int_{E_{\rm min}^\prime}^\infty dE^\prime
  f_N(E^\prime-q_0)[1-f_N(E^\prime)],
  \label{eq:tilde S}\\
 E_{\rm min}^\prime&=&-\mu_BBs^\prime
  +\frac{[q_0+q^2/2m-\mu_BB(s-s^\prime)]^2}{4q^2/2m}.
  \label{eq:E'_min}
\end{eqnarray}

From this point on, we show $\tilde S_{ss^\prime}=S_{ss^\prime}$ for
nondegenerate nucleons.
We first define the dimensionless variables
$x^\prime=(E^\prime-\mu_N)/T$ and $z=q_0/T$; the integral yields
\begin{eqnarray}
 \tilde S_{ss^\prime}(q_0,q)&=&
  \frac{m^2T}{2\pi q}\int_{x^\prime_{\rm min}}^\infty dx^\prime
  \left(\frac{1}{e^{x^\prime-z}+1}\right)
  \left(\frac{1}{1+e^{-x^\prime}}\right)\nonumber\\
  &=&\frac{m^2T}{2\pi q}\frac{1}{1-e^{-z}}
  \ln\left(\frac{1+e^{-x^\prime_{\rm min}+z}}
      {1+e^{-x^\prime_{\rm min}}}\right).
  \label{eq:tilde S after integration}
\end{eqnarray}
Expanding $x^\prime_{\rm min}$ to linear order in $B$ we find
\begin{eqnarray}
 x^\prime_{\rm min}&\simeq& x^\prime_0+\delta x^\prime,\nonumber\\
 x^\prime_0&=&\frac{(q_0+q^2/2m)^2}{4T(q^2/2m)}-\frac{\mu_N}{T},\nonumber\\
 \delta x^\prime&=&\frac{-\mu_BB}{2T}
  \left[\left(1+\frac{2mq_0}{q^2}\right)s
   +\left(1-\frac{2mq_0}{q^2}s^\prime\right)\right].
   \label{eq:expansion for x}
\end{eqnarray}
Note that $\delta x^\prime$ is identical to $\delta x$
[Eq. (\ref{eq:parameters})] which already appeared in deriving nucleon
response function.
For $\delta x\ll 1$, $\tilde S_{ss^\prime}$ can be written as a sum of
$\tilde S_0$, the zero field value, and $\delta\tilde S_{ss^\prime}$,
the correction due to the magnetic field, i.e.,
\begin{eqnarray}
 \tilde S_{ss^\prime}(q_0,q)&=&\tilde S_0(q_0,q)
  +\delta\tilde S_{ss^\prime}(q_0,q),\nonumber\\
 \tilde S_0(q_0,q)&=&\frac{m^2T}{2\pi q}
  \frac{1}{1-e^{-z}}\ln\left(\frac{1+e^{-x^\prime_0+z}}
		      {1+e^{-x^\prime_0}}\right),\nonumber\\
 \delta\tilde S_{ss^\prime}(q_0,q)&=&-\frac{m^2T}{2\pi q}
  \frac{\delta x}{\left(1+e^{x^\prime_0-z}\right)
  \left(1+e^{-x^\prime_0}\right)}.
  \label{eq:expansion for tilde S}
\end{eqnarray}
In the limit of nondegenerate nucleons with $\mu_N/T\ll -1$ and
$\exp(\mu_N/T)=(2\pi^3/m^3T^3)^{1/2}n$, we find
\begin{eqnarray}
 \lefteqn{\frac{1}{1-e^{-z}}\ln\left(\frac{1+e^{-x^\prime_0+z}}
		     {1+e^{-x^\prime_0}}\right)}\nonumber\\
 &\simeq&\frac{1}{\left(1+e^{x^\prime_0-z}\right)
  \left(1+e^{-x^\prime_0}\right)}\nonumber\\
 &\simeq&e^{-x^\prime_0+z}
  =\exp\left[\frac{\mu_N}{T}
	-\frac{(q_0-q^2/2m)^2}{4T(q^2/2m)}\right].
  \label{eq:approx}
\end{eqnarray}
This last form is the same as that of Eq. (B9) in Ref. \citep{Arras99b},
which is the $S_{ss^\prime}$ limit for nondegenerate nucleons,
guaranteeing the relation $\tilde S_{ss^\prime}=S_{ss^\prime}$ for
nondegenerate nucleon limit.

Hence, we obtain the expression for the momentum integral which includes
$\Vec P$ as follows:
\begin{eqnarray}
 \lefteqn{\int\frac{d^3p}{(2\pi)^3}\frac{d^3p^\prime}{(2\pi)^3}
  (2\pi)^4\delta^4(p+k-p^\prime-k^\prime)
  f_N(1-f_N^\prime)\Vec P}\nonumber\\
 &=&\frac{2m\Vec q}{q^2}
  \left[q_0-\mu_BB(s-s^\prime)\right]S_{ss^\prime}(q_0,q).
  \label{eq:final P integral}
\end{eqnarray}
The nuclear response function $S_{ss^\prime}(q_0,q)$ can be easily
expanded by $\epsilon$ using the same method given in
Ref. \citep{Arras99b}; leading to Eq. (\ref{eq:S_0 for nondegenerate}).
Thus, all the expressions of Eqs. (\ref{eq:A_0 for
nondegenerate})--(\ref{eq:delta C_WM for nondegenerate}) are also
obtained.

\begin{acknowledgments}
The author is supported by the Research Fellowships of the Japan Society
 for the Promotion of Science for Young Scientists.
\end{acknowledgments}

\bibliography{refs}

\end{document}